\documentclass[conference]{IEEEtran}
\usepackage{graphicx}
\usepackage{balance}

\begin{document}

\title{ChatEd: A Chatbot Leveraging ChatGPT for an Enhanced Learning Experience in Higher Education}

\author{\IEEEauthorblockN{Kevin Wang}
\IEEEauthorblockA{Department of Computer Science\\
University of British Columbia\\
Kelowna, BC, Canada, V1V 2Z3\\
wskksw@mail.ubc.ca}
\and
\IEEEauthorblockN{Jason Ramos}
\IEEEauthorblockA{Department of Computer Science\\
University of British Columbia\\
Kelowna, BC, Canada, V1V 2Z3\\
ramosjasonwork@gmail.com}
\and
\IEEEauthorblockN{Ramon Lawrence}
\IEEEauthorblockA{Department of Computer Science\\
University of British Columbia\\
Kelowna, BC, Canada, V1V 2Z3\\
ramon.lawrence@ubc.ca}
}

\maketitle

\begin{abstract}

With the rapid evolution of Natural Language Processing (NLP), Large Language Models (LLMs) like ChatGPT have emerged as powerful tools capable of transforming various sectors. Their vast knowledge base and dynamic interaction capabilities represent significant potential in improving education by operating as a personalized assistant. However, the possibility of generating incorrect, biased, or unhelpful answers are a key challenge to resolve when deploying LLMs in an education context. This work introduces an innovative architecture that combines the strengths of ChatGPT with a traditional information retrieval based chatbot framework to offer enhanced student support in higher education. Our empirical evaluations underscore the high promise of this approach.

\end{abstract}


\maketitle

\section{Introduction}

Large Language Models (LLMs) like ChatGPT present a new frontier in education with their abilities to transform education by providing personalized learning and generating and answering questions. However, there are known challenges including concerns about misinformation, a lack of domain-specific expertise, and potential biases \cite{michel2023challenges,kooli2023chatbots,kasneci2023chatgpt}. Accuracy and specificity are paramount in education.

Virtual assistants tailored for educational contexts have emerged. A virtual assistant named Jill Watson trained the chatbot on previous and ongoing Q\&A to provide customized assistance for courses \cite{goel2018jill}. Prior to LLMs, issues with deploying chatbots included significant overhead associated with training the models for every new course, a dependency on vast and accurate data, and their often limited conversational scope. Prior virtual assistants were confined to data they started with unlike the vast amount of data on which LLMs are trained on.

Given these challenges, we designed and implemented a new chatbot architecture for education called ChatEd, which is retrieval-based and integrated with a large language model such as ChatGPT. Retrieval-based models, traditionally employed by specifically trained chatbots, select the most appropriate document (chunk of information) from a predefined set of documents (usually derived from course materials) based on the user's input, which ensures validity and relevancy. Instead of returning only the reference document and answer, the document is sent to a LLM for generating the query response. This integration with the LLM introduces the benefits of human-like conversations, contextual understanding, and depth of conversation, while preserving the key benefits of traditional information retrieval by ensuring accuracy of responses and verified references to original source documents. The combination has better usability than traditional retrieval chatbots and improved accuracy and source verification compared to using only an LLM. Providing original sources allows students to locate the information. The key system features include:

\begin{itemize}
    \item \textbf{Enhanced Accuracy and Contextual Relevance}: By combining the power of NLP with a rich knowledge base, the system delivers contextually accurate responses. 
    
    \item \textbf{Conversational Memory}: The model is designed to remember and reference chat history, allowing for deeper and more meaningful interactions.
    
    \item \textbf{Ease of Integration}: A user-friendly interface allows seamless integration with existing Learning Management Systems (LMS) and leverages existing course materials to ensure accurate and educational context-specific answers. 
    
    \item \textbf{Streamlined and Easy Training}: Unlike previous chatbot approaches, there is no training required on the Q\&A data. The system leverages existing course materials for a more efficient chatbot building process.
\end{itemize}

The primary contribution is introducing and evaluating a new chatbot framework for educational use. We show that the system has a very high level of question answering ability through leveraging both a retrieval based framework and ChatGPT. A distinctive feature of our model is its scalability and ease of use for instructors. Instead of relying on traditional and labor-intensive methods of chatbot training, our model facilitates the direct input of existing course materials. This approach streamlines the setup process and ensures that the chatbot's knowledge base is aligned with the course content. 

A comprehensive series of evaluations is performed to assess the core framework of our proposed system, particularly its question-answering ability and context awareness. Evaluations are performed to measure if the question answers are helpful, relevant, accurate, and correct. Experiments show the ChatEd chatbot excels in all these criteria. ChatEd is compared with ChatGPT for answer quality and ability to support conversational interactions.

\section{Background}

\subsection{Large Language Models and Education}

LLMs such as ChatGPT provide novel opportunities in education \cite{michel2023challenges, kooli2023chatbots,LoRapidReview}. These systems can provide personalized learning and function as a research assistant and 24/7 online instructor \cite{michel2023challenges,kooli2023chatbots,kasneci2023chatgpt}. However, LLMs have also been shown to generate incorrect, misleading, and biased information and rarely provide correct references to original sources. Misinformation is a significant concern in education, including false information or biased information. Another problem is that they leverage general information and lack domain specific knowledge, which poses problems especially in education as students need specific guidance in fields that is relevant to their particular course and instructional context \cite{kooli2023chatbots}.

\subsection{Virtual Assistants in Higher Education}

There is prior research on developing virtual assistants (teaching agents) for courses. Jill Watson, made by Ashok {\em et al.} \cite{goel2018jill}, is one of the most renowned virtual assistants for educational questions. Jill Watson was originally designed for large MOOC courses using IBM Bluemix toolsuite.  Experimentation demonstrated that Jill Watson is effective in reducing teacher workload, promotes engagement, and is malleable \cite{goel2018jill}. Jill Watson is developed based on training of Q\&A and a feedback loop, and it works within a discussion forum to answer questions that it is confident in answering. Through continuous developing and exploration, Jill Watson's versions became easier to train, and required under 24 hours to develop a new chatbot for a syllabi \cite{goel2020ai}. 

Cunningham {\em et al.} \cite{cunningham2019review} covered the common chatbot frameworks used, such as Dialogflow, RASA, and Wit.ai. They described applications for chatbots in education such as a FAQ Chatbot, but also cite the drawbacks such as needing accurate data for training.

Kuhail {\em et al.} \cite{kuhail2023interacting} conducted a comprehensive review of chatbots in education including their limitations. A common problem with chatbots is their limited capabilities to expand beyond the initial data they are trained on. For educational purposes, instructors can discuss the problem and answer new related questions. This is hard to do unless the chatbot is provided with an extremely large amount of data. By deploying Jill Watson in a MOOC, student Q\&A was abundant and the cost of training a chatbot can be justified. Even so, the conversational ability is limited. The cost and challenges of training chatbots effectively blocks individual instructors from building specialized chatbots for their courses, as there is a bottleneck for both collecting the training data and the cost of training.

\subsection{Bridging the Gap}

The future of educational assistance lies in harnessing the power of LLMs while addressing their inherent shortcomings. We propose a novel system that marries the vast knowledge base of models like ChatGPT with the specific context provided by traditional virtual assistants. This synergy promises a chatbot that is both knowledgeable and specialized, offering students a comprehensive educational companion without the typical constraints of current models. It also offers the ability for instructors to develop and deploy chatbots for their own courses efficiently.

\section{System Design}

The ChatEd chatbot architecture is designed for easy deployment in courses by instructors and features a conversational, chat interface for students. The key requirements are:

\begin{itemize}
    \item {\bf Correctness} - chatbot must provide correct answers and communicate its limitations when a question is out of its knowledge scope.        
    \item {\bf Relevancy} - responses must be context-specific to the course material and educational context.
    
    \item {\bf Verifiable} - responses must be verifiable by students by providing specific source information and references used to generate the response.
    
\end{itemize}

\begin{figure}
    \centering
    \includegraphics[width=0.8\linewidth]{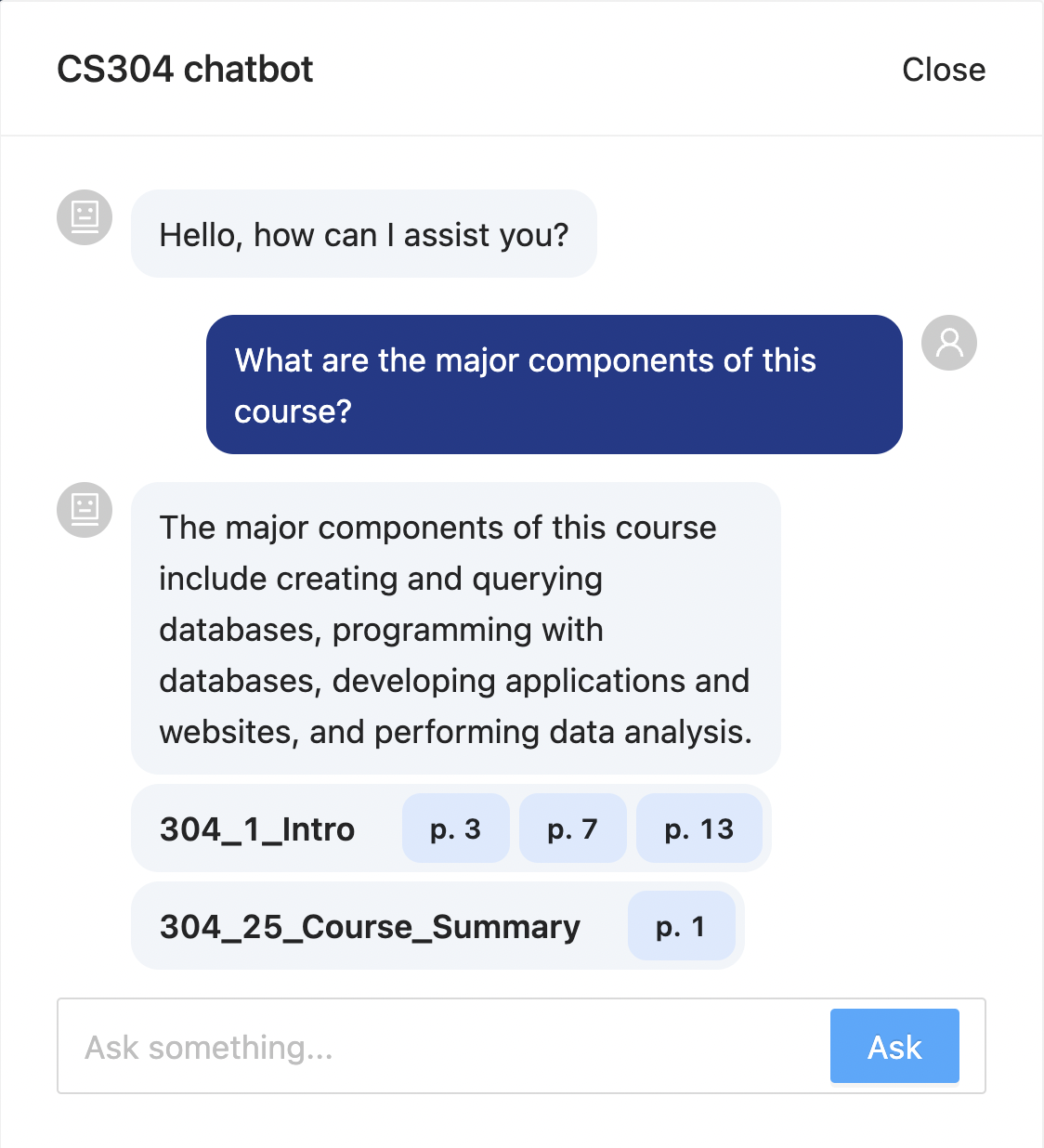}
    \caption{Chatbot User Interface and Example Conversation}
    \label{fig:chatbotui}
\end{figure}

Students interact with the chatbot similar to other chatbot systems by asking questions, receiving responses, and continuing a conversation with follow up questions and requests (see Figure \ref{fig:chatbotui}). The chatbot is a service that can be integrated with Learning Management Systems (LMSs), discussion and Q\&A forums, or other student interaction software. 

\begin{figure}
    \centering
    \includegraphics[width=1\linewidth]{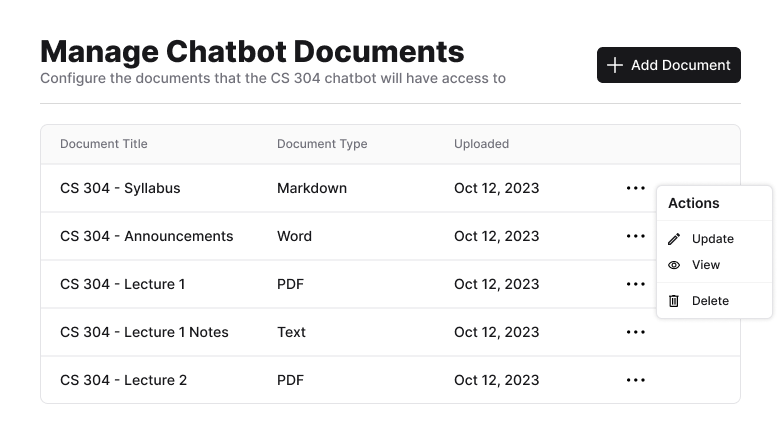}
    \caption{Instructor Interface}
    \label{fig:instructorInterface}
\end{figure}

For instructors, the system provides an efficient mechanism for training on source material including notes (PowerPoint, PDF), web sites, posts and announcements, and other textual content. These source materials are processed and summarized in a database for efficient retrieval and integration with ChatGPT. Importantly, this approach eliminates any requirements for training a model by the instructor and leaves the challenging data integration and conversational task to the LLM, such as ChatGPT. Source content is provided to the system by uploading files or providing URLs of source locations. The instructor interface is in Figure \ref{fig:instructorInterface}. 

\subsection{ChatEd Chatbot Architecture}

\begin{figure}
    \centering
    \includegraphics[width=0.9\linewidth]{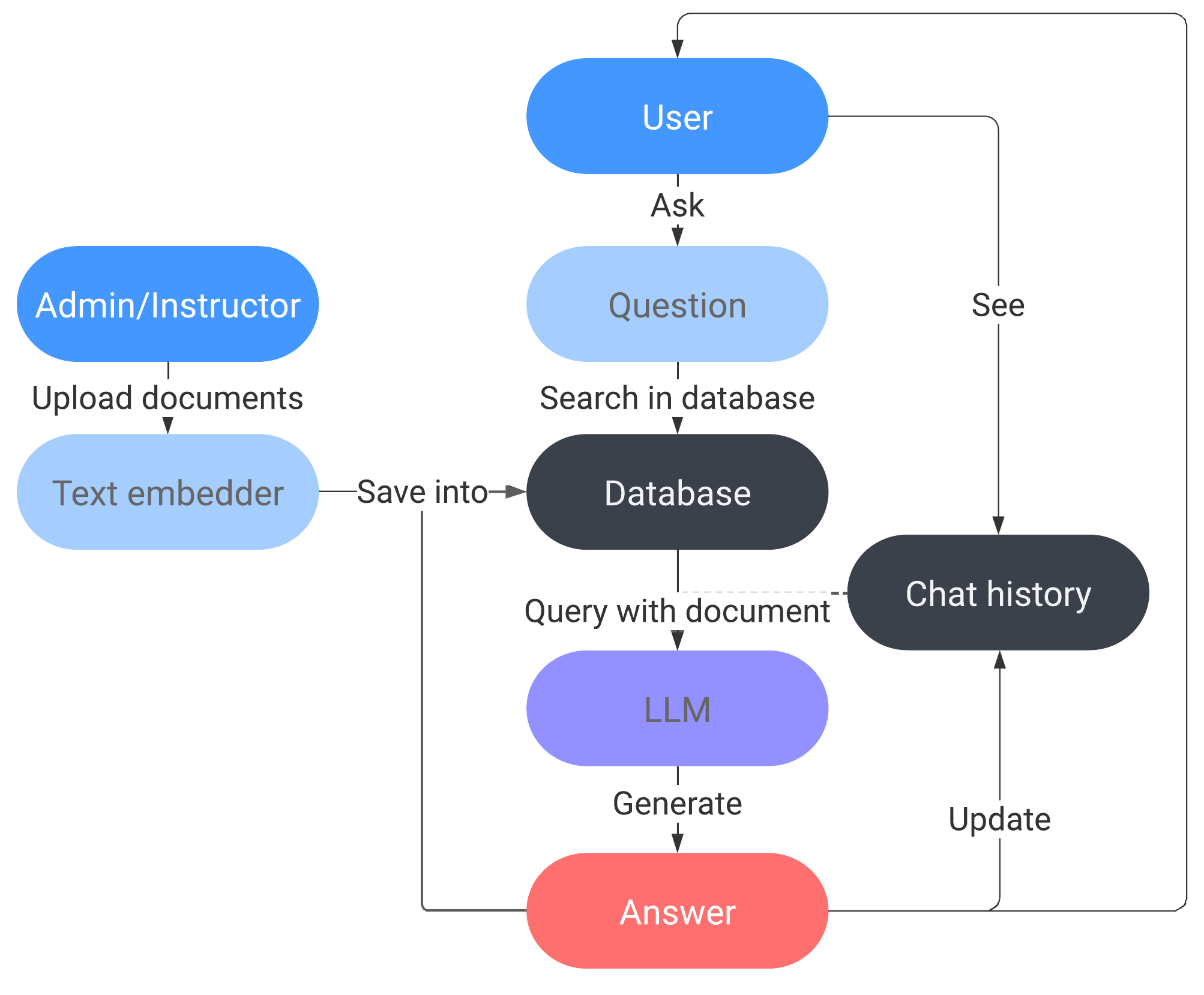}
    \caption{ChatEd Chatbot Architecture}
    \label{fig:framework}
\end{figure}

The unique aspect of the ChatEd architecture (Figure \ref{fig:framework}) is integrating an information retrieval system that stores and queries sources provided by the instructor with an LLM that provides the conversational support and general knowledge. 

\subsubsection{Context-Specific Database}

The first step is for instructors to provide their sources as documents or URLs. Each document is retrieved and indexed. 
For example, a PDF document containing a PowerPoint presentation is divided and indexed by slide and keywords captured on each indexed element. These instructor documents provide the source context for the chatbot that is specific to the current course. The database storing these documents is PostgreSQL, and similarity search is performed using Facebook AI Similarity Search library (Faiss\footnote{https://github.com/facebookresearch/faiss}). There is a separate database for each course or unique chatbot constructed.

\subsubsection{LLM Integration}

When a user poses a question, instead of sending the question directly to the LLM, which would respond using its generalized knowledge, the question is first used as a query in the database to determine similar indexed documents. Then, the question, indexed documents, and prior chat history are provided as a prompt to the LLM. This has the key advantages that the LLM has specific, contextualized knowledge to frame a response, and this response when provided back to the user can be augmented with document sources. The inclusion of chat history allows the LLM to maintain context, ensuring continuity in extended interactions. The LLM processes these inputs to generate a coherent and contextually relevant answer for the user. 

There is no local training required for the chatbot as the question answering and conversational aspects are provided by the LLM. The chatbot leverages the LangChain\footnote{https://www.langchain.com/} framework, which provides libraries that abstract the process of querying for relevant information and calling to Large Language Models (LLM). In our experiment, we used ChatGPT 3.5 turbo API.

\section{Methodology}

The ChatEd chatbot is evaluated for performance based on questions from two courses at a research-intensive university. The questions and answers were collected as part of a previous research study \cite{helpme} evaluating the types of questions asked during office hours and over email. The first course is CS304: Introduction to Databases which covers basic database knowledge such as SQL and database design. The second course is CS 404: Database Implementation which covers database internals and how to build a database system. The course materials for both courses are distributed on GitHub, including lecture slides, labs, and syllabi. The course materials were uploaded to the chatbot, and specific chatbots were created for each course.

The test data set consisted of a diverse variety of questions, including {\em general questions} such as `Can each table have multiple foreign keys?',  {\em domain-specific questions} such as `Are ``chunk" and ``block" interchangeable terms?', and {\em managerial questions} such as `Is the exam open-book?'. In total, there were 60 individual questions. 

The evaluation performed 3 different types of tests that used the question text, expected answer manually added according to lecture content, actual chatbot system response, and human evaluation of the chatbot response. The evaluation was performed by the instructor and teaching assistants of the courses.  The specific test types are outlined below. To benchmark the  system, a side-by-side comparison was conducted with ChatGPT, assessing the responses, accuracy, and relevance of both systems to identical questions.

\subsection{Question Answering Abilities}

ChatEd's question-answering ability was measured using the question set with responses, and three metrics were used in the evaluation:

\textbf{Relevance}: Evaluates the pertinence of ChatEd's responses to the question asked by the user. It assesses how well the response aligns with the specific query, ensuring that the answer directly addresses the main point of the question. 
\begin{enumerate}
    \item \textbf{Directness}: Does the response directly answer the user's question?
    \item \textbf{Conciseness}: Is the response free from unnecessary wordiness or details?
    \item \textbf{Contextual Awareness}: Does the response consider the previous questions or statements?
    \item  \textbf{Detail Tailoring}: Does the response address the nuances and specific details of the user's question?
    \item \textbf{Topic Alignment}: Does the response stay within the bounds of the subject matter?
\end{enumerate}

    \textbf{Accuracy}: Evaluates the extent to which ChatEd's responses provide factually correct information, particularly within the context of the specific topic or domain in question.
\begin{enumerate}
    \item \textbf{Factual Correctness}: Does the response provide information that is correct based on the documents provided?
    \item \textbf{Acknowledgment of Limitations}: Does the response transparently acknowledge any limitations in its knowledge base or ability to help?
    \item \textbf{Logical Consistency}: Does the response contain conflicting statements or information?
    \item \textbf{Completeness}: Is the response comprehensive and thorough, leaving no critical aspects of the user's query unaddressed?
    \item \textbf{Information Source}: Does the response use any information not contained in the documents provided?
\end{enumerate}

    \textbf{Helpfulness}: Evaluates how well ChatEd's responses address the user's question and how well it enhances the user's experience ensuring the overall utility of the response.
\begin{enumerate}
    \item \textbf{Clarity}: Is the response presented in a clear and understandable manner, avoiding jargon or overly complex language?
    \item \textbf{Empathy and Tone}: Is the response delivered in a friendly, respectful, and empathetic tone?
    \item \textbf{Follow-Up}: Is the response more detailed when the user asks for clarifications on previously asked questions?
    \item \textbf{Error Correction}: Does the response correct any inaccuracies or misconceptions in the user's question?
    \item \textbf{Source Citation}: Does the response provide the sources that it referenced?
\end{enumerate}

\subsection{Context Awareness Test Procedure}

The ability of ChatEd to understand, maintain, and switch contexts during interactions was tested using these steps:

\begin{enumerate}
\item \textbf{Establish a Context (A)}: Initiate with a context-setting question.
\item \textbf{Follow-up Without Explicit Context (B)}: Pose questions that implicitly rely on the established context.
\item \textbf{Deepen the Context (C)}: Delve deeper into the context with more specific questions.
\item \textbf{Break the Context (D)}: Interrupt the context sequence with an unrelated query.
\item \textbf{Revisit the Context (E)}: Return to the original context after the interruption.
\end{enumerate}

These steps were executed in various sequences to test the flexibility and memory of the chatbot. Responses were documented and analyzed to ascertain the ChatEd's proficiency in context-aware interactions. In each course, five distinct interactions were executed to evaluate consistency and reliability.

\section{Results}

\subsection{Question Answering Results}

\begin{figure}
    \centering
    \includegraphics[width=0.8\linewidth]{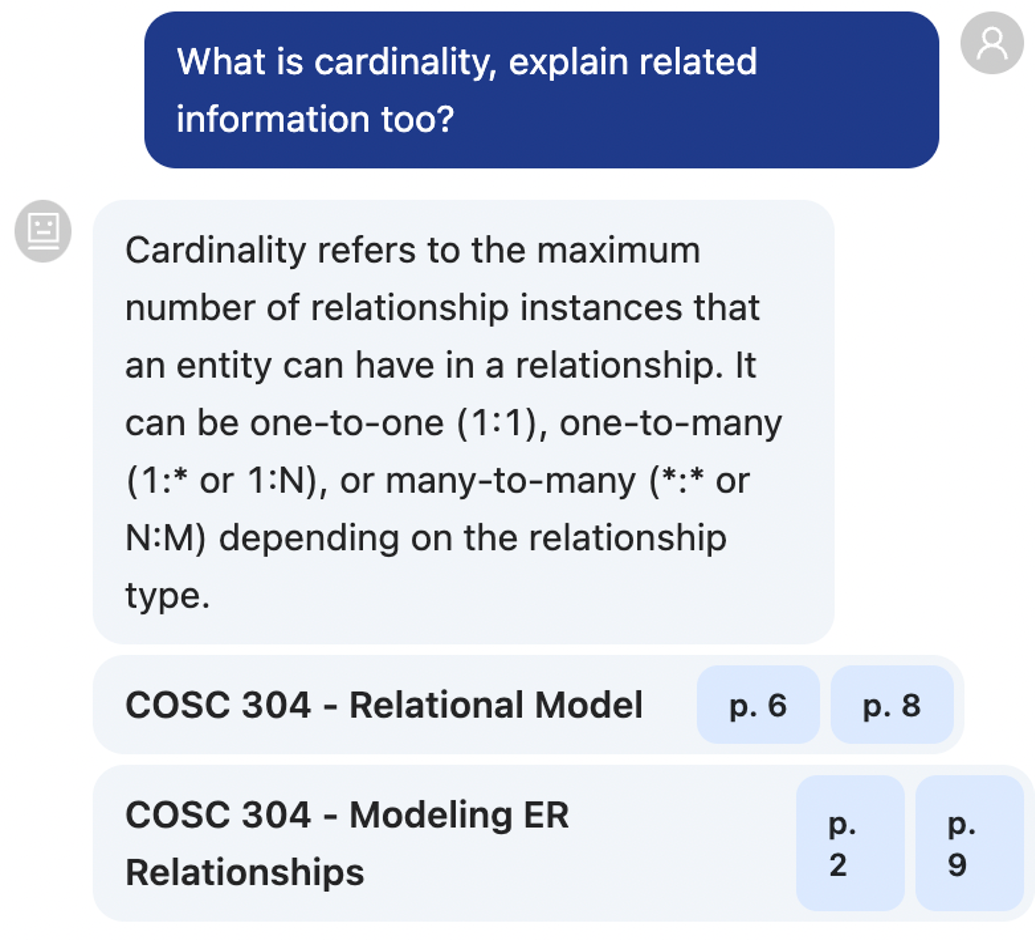}
    \caption{Example Interaction for Q\&A Test}
    \label{fig:qatest}
\end{figure}

ChatEd was evaluated using a sample of 20 questions from the question bank. Each question was posed to the system, and its answer was evaluated on a scale of 1 to 5, with 5 being the best, based on three criteria: relevancy, accuracy, and helpfulness. The questions were also directly posed to ChatGPT v3.5 for comparison. The results are in Table \ref{tbl:qaresults} and show the average score over all questions. An example test interaction is in Figure \ref{fig:qatest}.

\begin{table}[h!]
\centering
\begin{tabular}{|l|c|c|}
\hline
\textbf{Criteria} & \textbf{ChatEd Score} & \textbf{ChatGPT  Score} \\
\hline
Relevancy & 5& 4.4\\
\hline
Accuracy & 5& 4.4\\
\hline
Helpfulness & 4.5& 3.4\\
\hline
\end{tabular}
\caption{Average Scores Given by Evaluators}
\label{tbl:qaresults}
\end{table}

ChatEd had exceptional performance in the Q\&A test returning answers that were evaluated as highly relevant and accurate. The provided references to source material are helpful for students to gather more information on the subject. ChatEd answers questions regarding the course more accurately in the context of the course. 
ChatGPT scored lower for reasons summarized below.

\begin{itemize}
    \item ChatEd performs extremely well with managerial questions that LLMs cannot answer, such as policies of the course. The answers are specialized and contain information not available on the training data of ChatGPT. 
    
    \item ChatEd also supplies more concise responses compared to ChatGPT, which often provides more verbose responses with more unrelated content.
    
    \item Comments by teaching assistant reviewers highlight ChatEd's ability to return more precise answers that are specific to the course content. For example, ChatGPT's answer to the question `What are some errors when retrieving results? Explain them.' had 12 exhaustive bullet points while ChatEd provided the 5 common errors when retrieving database results covered in class.
\end{itemize}

\subsection{Conversational Depth}

\begin{figure}
    \centering
    \includegraphics[width=0.5\linewidth]{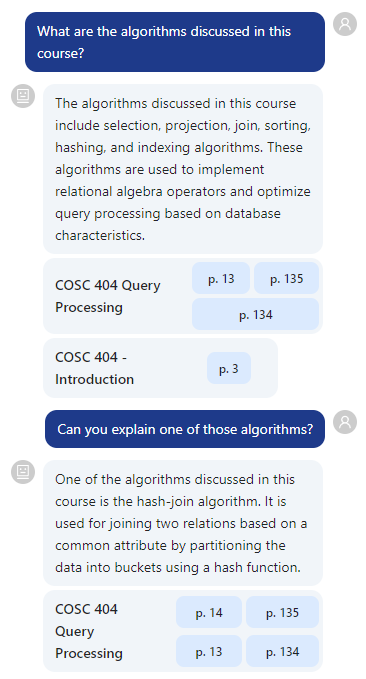}
    \caption{Example of Switching Contexts in a Conversation}
    \label{fig:convtest}
\end{figure}

The test for conversational depth evaluated ChatEd's ability to continue to provide relevant information as follow-up questions are asked by the user and adeptly respond when the user changes the question context during the conversation. The conversations were evaluated on a 5-point scale. The results are in Table \ref{tbl:convresults}. An example test interaction is in Figure \ref{fig:convtest}. ChatEd displayed strong results in all criteria especially accuracy, highlighting its capabilities to relate to chat history. 

\begin{table}[h!]
\centering
\begin{tabular}{|l|c|}
\hline
\textbf{Criteria} & \textbf{ChatEd Avg. Score} \\
\hline
Relevancy & 4.19\\
\hline
Accuracy & 5.00\\
\hline
Helpfulness & 4.06\\
\hline
\end{tabular}
\caption{Average Scores Given by Evaluators}
 \label{tbl:convresults}
\end{table}

Although it had a strong ability to be aware of the context of the conversation, there is still room for improvement. For example, when asked ``Why do we need this?" as a follow-up question to ``What is a hash function?", ChatEd responded with the importance of a different topic. Although the answer was factually correct, ChatEd failed to understand what ``this" was referring to. This resulted in an unhelpful and irrelevant response. We speculate that this is because the documents provided to ChatEd contained information about the importance of other topics. As it has access to both the history of the conversation and the relevant documents, ChatEd may have confused the documents with the original question.
Further, ChatEd is unable to consistently understand questions that refer to other questions that are too far back in the conversation. Without the context of the previous question, this results in ChatEd providing irrelevant or incomplete information. Overall, ChatEd shows promising potential as a conversational chatbot and will iteratively improve over time.

\section{Discussion}

Prior research on educational chatbots has shown good success and educational impact, especially in large classes. A key challenge was the cost of training chatbots on course-specific material, and their ability to adapt to new questions. The major advantage of ChatEd is that it requires no training by the instructor, who must only provide the source documents to the system for indexing. By using the source material as a prompt to a LLM rather than training the chatbot itself, the system leverages the general knowledge and conversational ability of the LLM while preserving local educational context and source references. Students receive relevant and contextually appropriate answers, enhancing their learning experience.

The improvement of ChatEd compared to ChatGPT depends on the source material provided. ChatGPT does a good job answering general questions that have accepted answers available on the Internet. It has weaker performance on questions that are on course specific content or less commonly known, and especially on questions that involve the management of the course such as deadlines, policies, and administrative factors.

\section{Conclusion and Future Work}

This work developed and evaluated a novel architecture for leveraging a LLM augmented with localized information for particular courses. By removing the need for local training of the chatbot, instructors can easily utilize the system in their own courses simply by providing the relevant materials. The evaluation performed on actual student questions showed high performance in the criteria of relevancy, accuracy, and helpfulness with performance improving upon ChatGPT that does not have the localized context information.

Future work will evaluate the system on questions in more courses, improve the question answering ability, and make the system available to instructors for use.

\bibliographystyle{IEEEtran}
\balance
\bibliography{ref.bib}

\section{Appendix}
Table \ref{tbl:interaction_ca3} shows a set of interactions with ChatEd.
Table \ref{tbl:convresults_selected} contains 5 sample questions used to test ChatEd on single Q\&As.

\begin{table*}
\centering
\begin{tabular}{|c|p{2in}|p{3.5in}|c|}
\hline
\textbf{Number} & \textbf{Question} & \textbf{Actual Response} \\
\hline
1 & What are transactions? & Transactions are atomic programs that execute on a database and maintain the consistency of the database. They ensure that the database remains in a consistent state before and after their execution.
COSC 404 - Transactions p. 4 p. 56
COSC 404 - Concurrency p. 62
COSC 404 - Scaling and Distribution p. 57  \\
\hline
2 & Can you explain what a consistent state is? & A consistent state in a database refers to a state where the data satisfies all the constraints specified in the database schema. It means that the data in the database is in accordance with the rules or predicates defined for it.
COSC 404 - Transactions p. 5 p. 58 p. 6
COSC 404 - Scaling and Distribution p. 15  \\
\hline
3 & How does a transaction do that? & A transaction ensures consistency by preserving the consistency of the database. It executes as an atomic program, meaning it either completes all its operations or none at all, ensuring that the database remains in a consistent state throughout the transaction.
COSC 404 - Transactions p. 4 p. 3 p. 10
COSC 404 - Concurrency p. 62  \\
\hline
4 & What's an index? & An index is a data structure that allows for fast lookup of records in a file. It can also help retrieve records in sorted order. Indexing is important for file systems and databases as it improves the efficiency of queries by minimizing the number of blocks accessed.
COSC 404 Indexing p. 2 p. 50 p. 4 p. 51  \\
\hline
5 & Can you further explain my first question? & An index is a data structure that allows for quick retrieval of records in a file. It improves query efficiency by minimizing the number of blocks accessed. Without secondary indexes, answering the query would require scanning the entire file, resulting in slower performance. With only one secondary index, the query could still be answered efficiently by retrieving the buckets for the indexed attribute and comparing the records.
COSC 404 Indexing p. 27
COSC 404 - Query Optimization p. 14 p. 10
COSC 404 - Transactions p. 16  \\
\hline
\end{tabular}
\caption{Sample interaction 
(rated 4/5/4 for relevance/accuracy/helpfulness)}
\label{tbl:interaction_ca3}
\end{table*}

\begin{table*}[h!]
\centering
\begin{tabular}{|p{1.5in}|p{2in}|c|p{2in}|c|}
\hline
\textbf{Question} & \textbf{ChatEd Response} & \textbf{Rating} & \textbf{ChatGPT Response} & \textbf{Rating} \\
\hline
What are the major components of this course? & The major components of this course include creating, querying, and programming with databases to develop applications, web sites, and perform data analysis. Additionally, the course covers database techniques and software including relational and NoSQL databases, SQL, JSON, and XML.\newline
COSC 304 - Course Introduction p. 3 p. 7 p. 13
COSC 304 - Course Summary and COSC 404 p. 1 & 5 / 5 / 4 & I'm sorry, but I need more specific information to provide you with an accurate answer. Could you please clarify which course you are referring to or provide more details about the subject or context of the course? Once I have more information, I'd be happy to help you identify the major components of that course. & 1 / 1 / 1 \\
\hline
Are ``chunk" and ``block" interchangeable terms?  & 
No, ``chunk" and ``block" are not interchangeable terms. In the context of the resources provided, a ``block" refers to a chunk of memory that can be read or written, while a ``chunk" is not specifically defined in the given documents.\newline
404\_2\_Storage p. 33 p. 64 p. 60
404\_12\_Recovery p. 8 & 5 / 5 / 5 & ``Chunk" and ``block" can be used interchangeably in some contexts, but they are not universally synonymous. The specific meaning of each term can vary depending on the domain or context. Here are a few examples: Memory and Cognitive Science: ``Chunking" is a process where individual pieces of information are bound together into a meaningful whole. ``Block" doesn't have the same... (word count: 226) & 1 / 1 / 1 \\
\hline
When are the midterm exams? & The midterm exams are on October 10th and November 2nd in class.\newline
Syllabus p. 7 p. 2 p. 5 p. 4 & 5 / 5 / 5 & I'm sorry, but I don't have access to your specific course schedule or information about when your midterm exams are. To find out the exact date and time of your midterm exams, you should: Check your course syllabus: The syllabus provided by your instructor at the beginning of the course typically includes information about the exam dates. Visit the course... (word count: 203) & 5 / 5 / 1 \\
\hline
How do I model a domain (explained in English) in an ER diagram? & To model a domain in an ER diagram, you need to identify the entities involved in the domain and their relationships. Represent each entity  as a rectangle and connect them with lines to represent the relationships between them.\newline
COSC 304 - Modeling ER Relationships p. 29
COSC 304 - Database Design and ER Modeling p. 32 p. 1
COSC 304 - ER to Relational Mapping p. 2 & 5 / 5 / 5 & Modeling a domain in an Entity-Relationship (ER) diagram involves representing the key entities, attributes, and relationships within a specific area of interest or problem domain. Here's a step-by-step guide on how to do this in simple English: Identify the Main Entities: Start by identifying the main entities or objects within the domain. These are the things you want to represent... (word count: 424) & 5 / 5 / 5 \\
\hline
Can each table have multiple foreign keys? & Yes, each table can have multiple foreign keys. COSC 304 - Relational Model p. 18 p. 21 \newline
COSC 304 - SQL DDL p. 17 p. 18 & 5 / 5 / 5 & Yes, a database table can have multiple foreign keys. Foreign keys are used to establish relationships between tables in a relational database. Each foreign key in a table references a primary key in another table, creating a link between the two tables. These relationships help maintain data integrity and allow for the retrieval of related information. Here are a few... (word count: 251) & 5 / 5 / 4 \\
\hline
\end{tabular}
\caption{Sample Questions from the Question Set}
\label{tbl:convresults_selected}
\end{table*}

\end{document}